# Ultra-low Frequency Acoustic Luneburg Lens

**Liuxian Zhao[1], Xuxu Zhuang[1], Hao Guo[1], Chuanxing Bi[1], Zhaoyong Sun[2,*]**

[1]Institute of Sound and Vibration Research, Hefei University of Technology, 193 Tunxi Road, Hefei 230009, China

[2]Beijing Institute of Graphic Communication,1 Xinghua Avenue (Band 2), Beijing, 102600, China

*Author to whom correspondence should be addressed: sunzhaoyong@bigc.edu.cn

**ABSTRACT**

In this paper, a novel structural Luneburg lens with local resonators is proposed. This lens allows for the realization of subwavelength focusing in low frequency range. The lens is achieved by graded refractive index from the lens's centre to the outer surface. Numerical simulations are conducted to obtain data on wave propagation waveform, maximum displacement amplitude, and full width at half maximum (FWHM) of the lens's focal region. The results show that a broadband frequency range can be achieved for subwavelength focusing. This provides a straightforward and adaptable method for designing the structural Luneburg lens for numerous applications.

# 1. Introduction

Gradient index (GRIN) lenses have garnered significant interest in recent years for their ability to operate and control the propagation of optical and acoustic waves [1-6]. Such lenses involve spatially varying refractive indices, which can be achieved by locally adjusting the geometric parameters of subwavelength unit cells. The effective refractive index of a GRIN lens is contingent upon the filling ratio of scatterers in the unit cell [7, 8]. Lin *et al* [9] firstly demonstrated the GRIN lens for structural waves. The manipulation of the propagation path of structural waves, in this case, was achieved through customizing the filling ratio of unit cells. For the propagation of structural waves in thin plates, it was proved that the effective velocity is linearly related to the plate thickness [10-13].

One of the intriguing applications of wave propagation in plates is vibration-based energy harvesting [14-19]. Researchers have explored the use of GRIN lenses to focus plane structural waves for optimized energy collection. Environmental vibrations, typically in the far field, can be approximated as plane wave excitations and energy harvested at the lens' focus [20, 21]. For instance, Tol *et al.* [22] introduced a GRIN lens that consisted of cylinder arrays with varying height and reported up to triple the power output when a piezoelectric transducer was placed at the lens focus. Similarly, Zhao *et al.* [23] devised a GRIN structural Luneburg lens using continuous variation of the plate thickness and displayed a nine-fold increase in energy harvesting at the lens's focal point. In a different study, Zhao *et al.* [24] examined a planar GRIN lens with a fluctuating thickness. The lens exhibited an impressive displacement amplitude at the focus, measuring approximately 30 to 40 times greater than that of the incident wave. Notably, all the aforementioned techniques are effective specifically with structural waves at high frequencies.

In GRIN structures, the common methods for generating graded refractive index are Bragg scattering and local resonances [25, 26]. In the high-frequency range, Bragg scattering is generally used for high frequencies where the wavelength is in the same scale as the unit cell size [12]. In addition, graded refractive index can also be obtained based on effective material properties, such as graded pentamode metamaterials for cloaking or metasurface [27, 28]. Metamaterial devices based on such design is, in principle, valid for middle frequency ranges. On the other hand, at much lower frequencies than Bragg scattering, subwavelength focusing can be generated using local resonance. Control of the resonant-type structural Luneburg lens due to local resonance mechanism is easily possible. In order to adjust the variation of refractive index, the dimensions of the locally resonant unit cell can be changed. In this paper, motivated by the concept of locally resonant GRIN, we explore a broadband resonant type structural Luneburg lens, which enables the realization of low frequency subwavelength focusing.

## 2. Design of the unit cells

The refractive index distribution of the Luneburg lens can be expressed as follows:

$$n(r) = \sqrt{2 - (r/R)^2},$$

where *r* represents the radial distance from the central point, and *R* is the lens's radius (*R* = 0.2 m in this study).

It is evident that the refractive index exhibits a centrally symmetric distribution. The value reaches its maximum of 1.414 at the center of the lens, while it decreases to a minimum of 1 at the edge of the lens, as depicted in Figure 1.

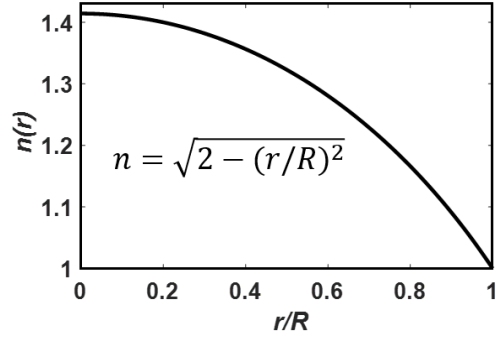

Figure 1: The distribution of refractive index of acoustic Luneburg lens.

For flexural waves, assuming the refractive index of the background plate as 1, the acoustic Luneburg lens achieves impedance matching with the plate. In this study, rather than achieving a gradient refractive index by adjusting the plate's thickness, we regulate the flexural waves by designing a local resonance structure. By employing this designed local resonance structure, a flexural wave Luneburg lens is designed and demonstrates the excellent functionality. Figure 2(a) displays a typical resonant unit cell, which is composed of a hexagonal plate with a circular hole and a resonant pillar, connected by three slender beams. This local resonance structure can be likened to a spring-oscillator system, with pillars serving as oscillators and slender beams as springs. The resonance frequency relates to the mass of the pillars and the stiffness of the beams. In our study, we maintain constant pillar diameter and beam cross-section, while adjusting the pillar height (referred to as $h_0$) and beam length (referred to as $R_0$) to control the effective wave velocity within the local resonance structure. The pillar height mainly controls the mass of the resonators and the beam length determines the coupling stiffness between the resonators and the hexagonal frames. Therefore, these two parameters, one for controlling mass and the other for stiffness, can well adjust the effective wave velocity within the local resonance structure.

The unit cell material is aluminium with a density of $\rho = 2700$ kg/m³, Young's modulus $E = 70$ GPa, and Poisson's ratio $v = 0.33$. The unit cell's dimensions are $a_0 = 20$ mm, $r_0 = 5$ mm, $w_0 = 1$ mm, and $d_0 = 2$ mm. The refractive index is a function of the parameters $h_0$ and $R_0$, derived from the slope of the dispersion curve calculated in the first Brillouin zone of the hexagonal unit cell, as illustrated in Figure 2(b) and (c). Thus, the refractive index is a function of parameters $h_0$ and $R_0$, determined based on the dispersion curve's slope. As an example, the $h_0$ and $R_0$ in Figure 2(a) are 8 mm and 15 mm, correspondingly. The dispersive curves of the structure are shown in Figure 2(c) where the first brillouin zone is demonstrated in Figure 2(b). According the dispersive curves, the first curve represents the flexural wave mode in which the effective velocity can be obtained by the formula of $c_F = 2\pi f/k$ with $k$ as the wave number.

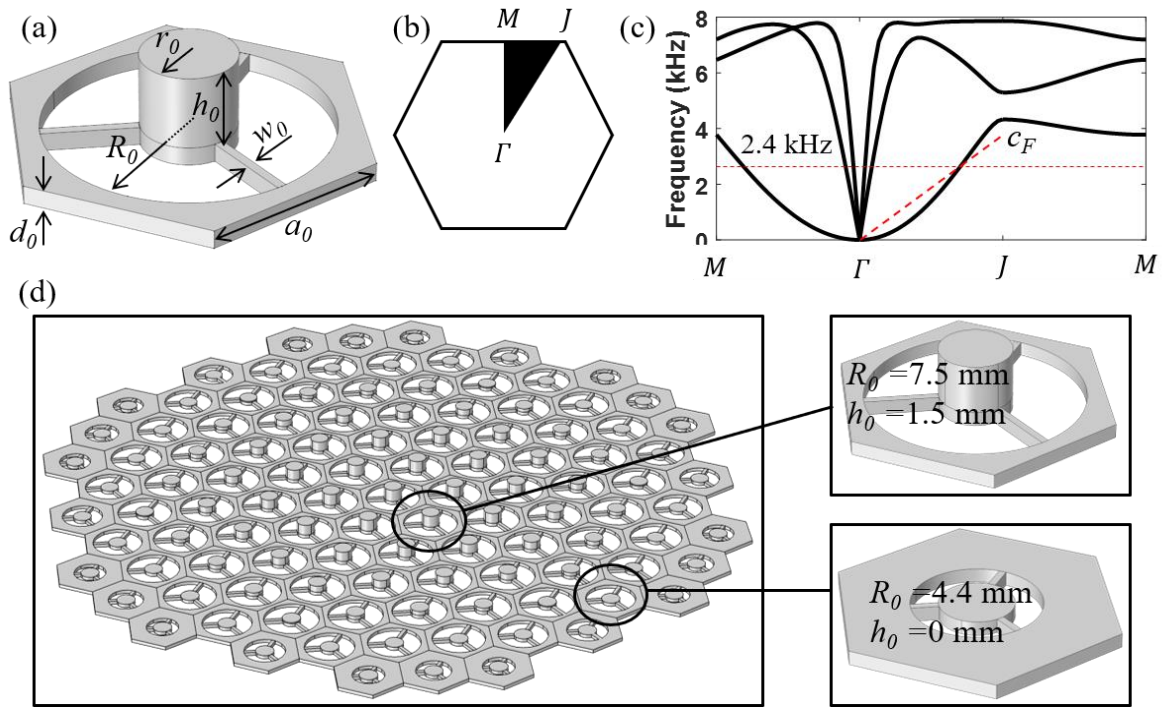

Figure 2: Design principles of resonant type structural Luneburg lens. (a) The hexagonal unit cell. (b)-(c) The first brillouin zone of the hexagonal unit cell and its dispersion curve. (d) The designed resonant type acoustic Luneburg lens.

In summary, the refractive index of the primitive cell can be determined by adjusting its geometric parameters, $R_0$ and $h_0$. Subsequently, an optimization algorithm is applied to identify the primitive cell that meets the desired refractive index distribution. The designed lens operates at a frequency of $f$ = 2.4 kHz and consists of 109-unit cells, as illustrated in Figure 2(d). Detailed geometric parameters are available in the Supplementary Materials.

## 3. Numerical Studies

The performance of the designed flexural wave lens was assessed through simulations using the finite element solver COMSOL Multiphysics, as depicted in Fig.3(a). The lens is coupled with a flat plate, which serves as the background medium. It matches the thickness of the lens's hexagonal base and shares the same material. The plate's dimensions are 800 mm × 600 mm × 2 mm. To minimize reflections from the plate's edges, Perfectly Matched Layers (PML) were employed (40 mm wide). The excitation signal used is a harmonic signal with a central frequency of 2.4 kHz, indicated by the red line in Figure 3(a). Figure 3(b) illustrates the displacement field distribution, revealing that the incident plane wave from the left side experiences refraction through the lens, resulting in noticeable convergence.

To evaluate the subwavelength focusing performance, we extracted displacements along the circumferential direction of the lens (as denoted by the black dotted line in Figure 3(b)) and depicted them in Figure 3(c). It is clearly observable that the displacement at $\theta = 0°$ surpasses that at other positions significantly. The calculation of the Full Width at Half Maximum (FWHM) results in FWHM = 0.56λ, which further solidifies the subwavelength attributes of the resonant-type structural Luneburg lens.

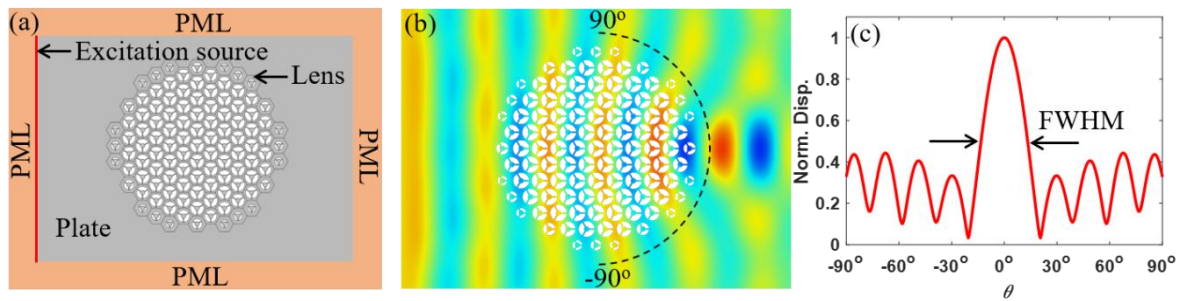

Figure 3: (a) The numerical model of thin plate structure with structural Luneburg lens. (b) Numerical simulation result. (c) The displacement amplitude along the circumferential direction of the lens.

Despite the lens's unit cells being designed to operate at 2.4 kHz, the dispersion curves, illustrated in Figure 2(c), indicate that the structure can function effectively across a wide range of frequencies. To substantiate this, frequency response tests were conducted using COMSOL software at various frequencies, including $f = 2$ kHz, $f = 2.4$ kHz, $f = 2.8$ kHz, $f = 3.2$ kHz, $f = 3.6$ kHz, and $f = 4$ kHz. Figure 4 presents the outcomes of the finite element simulation, vividly demonstrating the lens's capability for efficient wave focusing, concentrating plane waves at specific locations within a broadband frequency ranges.

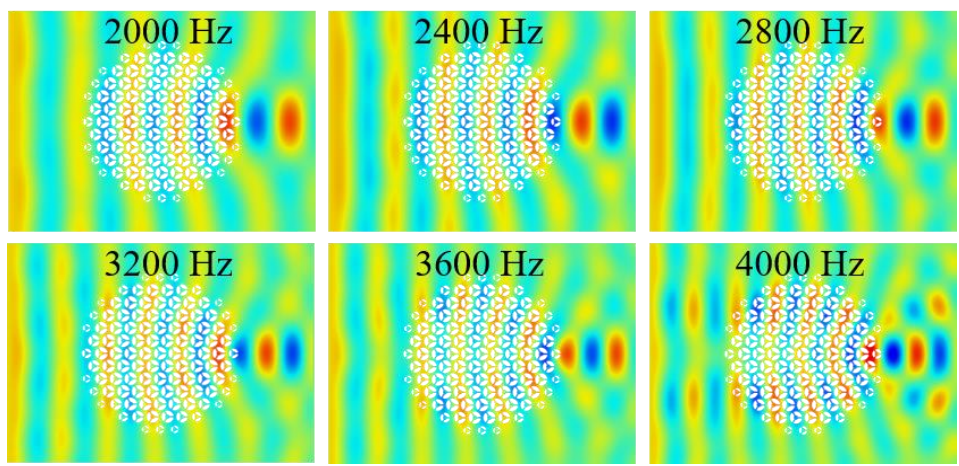

Figure 4: Numerical simulations at different frequencies.

In addition, the FWHM of the wave focusing are obtained based on the amplitude along the circumferential direction of the lens at the frequencies of $f = 2$ kHz, $f = 2.4$ kHz, $f = 2.8$ kHz, $f = 3.2$ kHz, $f = 3.6$ kHz, and $f = 4$ kHz, as shown in Figure 5. It's evident that the obtained FWHM fall within the subwavelength scale across a broad frequency spectrum. These findings underscore the ability of the resonant-type structural Luneburg lens to achieve subwavelength-scale focusing.

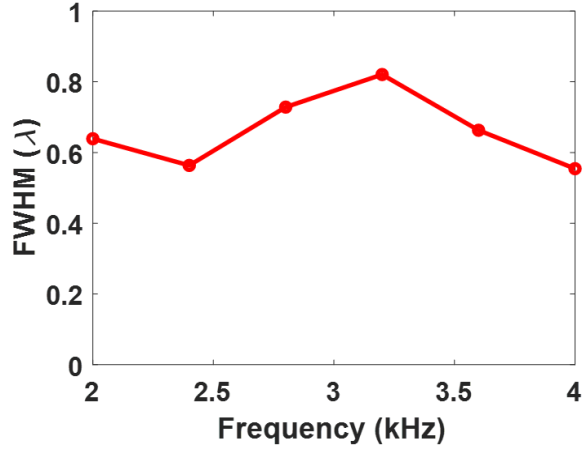

Figure 5: The FWHM obtained at different frequencies.

## 4. Conclusions

In conclusion, we introduce a resonant-type structure to design the Luneburg lens, in which the unit cell is composed of a hexagonal structure with a circular hole, a resonant pillar, and three slender beams to connect the hexagonal structure and pillar, functioning as local resonators. This lens configuration comprises 109 unit cells. We achieve effective refractive indices by fine-tuning the geometric parameters and retrieving cell bandwidths. The proposed lens design offers broadband, subwavelength focusing of bending waves at low frequencies. Numerical investigations confirm its subwavelength focusing capabilities, with a FWHM of approximately $0.5\lambda$ in the frequency range of 2-4 kHz. The implications of this research are expected to extend to various vibration-based energy harvesting techniques.

**Acknowledgments**

This work was supported by Anhui Provincial Natural Science Foundation (Grant No. JZ2023AKZR0583), the Fundamental Research Funds for the Central Universities (Grant No. JZ2023HGTB0215), and the BIGC Projects (Grant No. BIGC Ed202206, 27170123007).

**Conflict of Interest**

The authors declare no conflict of interest.

**Data Access Statement**

The data that support the findings of this study are available from the corresponding author upon reasonable request.

**Reference**

1. Tol, S., F.L. Degertekin, and A. Erturk, *Gradient-index phononic crystal lens-based enhancement of elastic wave energy harvesting.* Applied Physics Letters, 2016. **109**(6): p. 063902.
2. Liu, W., et al., *Manipulating light trace in a gradient-refractive-index medium: a Lagrangian optics method.* Optics Express, 2019. **27**(4): p. 4714-4726.
3. Jin, Y., B. Djafari-Rouhani, and D. Torrent, *Gradient index phononic crystals and metamaterials.* Nanophotonics, 2019. **8**: p. 685-701.
4. Park, J., et al., *Double-Focusing Gradient-Index Lens with Elastic Bragg Mirror for Highly Efficient Energy Harvesting.* Nanomaterials, 2022. **12**(6).
5. Zhao, L., et al., *Acoustic beam splitter based on acoustic metamaterial Luneburg lens.* Physics Letters A, 2023. **472**: p. 128815.
6. Zhao, L., et al., *A review of acoustic Luneburg lens: Physics and applications.* Mechanical Systems and Signal Processing, 2023. **199**: p. 110468.
7. Titovich, A.S., A.N. Norris, and M.R. Haberman, *A high transmission broadband gradient index lens using elastic shell acoustic metamaterial elements.* The Journal of the Acoustical Society of America, 2016. **139**(6): p. 3357-3364.
8. Ma, F., et al., *An underwater planar lens for broadband acoustic concentrator.* Applied Physics Letters, 2022. **120**(12): p. 121701.
9. Lin, S.-C.S., et al., *Gradient-index phononic crystals.* Physical Review B, 2009. **79**(9): p. 094302.
10. Climente, A., D. Torrent, and J. Sánchez-Dehesa, *Gradient index lenses for flexural waves based on thickness variations.* Applied Physics Letters, 2014. **105**(6): p. 064101.
11. Jin, Y., et al., *Gradient Index Devices for the Full Control of Elastic Waves in Plates.* Scientific Reports, 2016. **6**(1): p. 24437.
12. Jin, Y., et al., *Simultaneous control of the S0 and A0 Lamb modes by graded phononic crystal plates.* Journal of Applied Physics, 2015. **117**(24): p. 244904.
13. Zhao, L., S.C. Conlon, and F. Semperlotti, *Broadband energy harvesting using acoustic black hole structural tailoring.* Smart Materials and Structures, 2014. **23**(6): p. 065021.
14. Wang, J., et al., *Perspectives in flow-induced vibration energy harvesting.* Applied Physics Letters, 2021. **119**(10): p. 100502.
15. Aridogan, U., I. Basdogan, and A. Erturk, *Random vibration energy harvesting on thin plates using multiple piezopatches.* Journal of Intelligent Material Systems and Structures, 2016. **27**(20): p. 2744-2756.
16. Liang, H., G. Hao, and O.Z. Olszewski, *A review on vibration-based piezoelectric energy harvesting from the aspect of compliant mechanisms.* Sensors and Actuators A: Physical, 2021. **331**: p. 112743.

17. Hegendörfer, A., P. Steinmann, and J. Mergheim, *Nonlinear finite element system simulation of piezoelectric vibration-based energy harvesters.* Journal of Intelligent Material Systems and Structures, 2021. **33**(10): p. 1292-1307.
18. Hou, W., et al. *Prospects and Challenges of Flexible Stretchable Electrodes for Electronics*. Coatings, 2022. **12**, DOI: 10.3390/coatings12050558.
19. Liao, Q., et al., *Solid-phase sintering and vapor-liquid-solid growth of BP@MgO quantum dot crystals with a high piezoelectric response.* Journal of Advanced Ceramics, 2022. **11**(11): p. 1725-1734.
20. Zhao, L., et al., *Super-resolution imaging based on modified Maxwell's fish-eye lens.* Mechanical Systems and Signal Processing, 2024. **211**: p. 111232.
21. Zhao, L., C. Bi, and M. Yu, *Structural lens for broadband triple focusing and three-beam splitting of flexural waves.* International Journal of Mechanical Sciences, 2023. **240**: p. 107907.
22. Tol, S., F.L. Degertekin, and A. Erturk, *3D-printed phononic crystal lens for elastic wave focusing and energy harvesting.* Additive Manufacturing, 2019. **29**: p. 100780.
23. Zhao, L., C. Lai, and M. Yu, *Modified structural Luneburg lens for broadband focusing and collimation.* Mechanical Systems and Signal Processing, 2020. **144**: p. 106868.
24. Zhao, J., et al., *Broadband sub-diffraction and ultra-high energy density focusing of elastic waves in planar gradient-index lenses.* Journal of the Mechanics and Physics of Solids, 2021. **150**: p. 104357.
25. Wu, Z., et al., *Band-gap property of a novel elastic metamaterial beam with X-shaped local resonators.* Mechanical Systems and Signal Processing, 2019. **134**: p. 106357.
26. Wang, G., et al., *Enhancement of the vibration attenuation characteristics in local resonance metamaterial beams: Theory and experiment.* Mechanical Systems and Signal Processing, 2023. **188**: p. 110036.
27. Chen, Y. and G. Hu, *Broadband and High-Transmission Metasurface for Converting Underwater Cylindrical Waves to Plane Waves.* Physical Review Applied, 2019. **12**(4): p. 044046.
28. Chen, Y., et al., *Broadband solid cloak for underwater acoustics.* Physical Review B, 2017. **95**(18): p. 180104.

# Supplemental Materials of "Ultra-low Frequency Acoustic Luneburg Lens"

**Liuxian Zhao[1], Xuxu Zhuang[1], Hao Guo[1], Chuanxing Bi[1], Zhaoyong Sun[2]**

[1]Institute of Sound and Vibration Research, Hefei University of Technology, 193 Tunxi Road, Hefei 230009, China

[2]Beijing Institute of Graphic Communication,1 Xinghua Avenue (Band 2), Beijing, 102600, China

*Author to whom correspondence should be addressed: sunzhaoyong@bigc.edu.cn

Keywords: Structural Luneburg Lens; Subwavelength Focusing; Local Resonator

The lens consists of a total of 109-unit cells. Considering symmetry, there are only 35 independent unit cells with distinct geometric parameters. We begin by discretizing the lens and then calculating the refractive index distribution after discretization. Utilizing a band retrieval method, we select the 35 independent unit cells that meet our requirements. The geometric parameters are listed in Table 1, where X and Y are the coordinates of the unit cell center, measured in units of "$a$", and the coordinate origin is at the center of the lens. "$n$" represents the respective effective refractive index of the unit cell, while "$(h_0, R_0)$" denote the geometric parameters of the unit cell.

| No | $X(a)$ | $Y(a)$ | $n$ | $(h_0, R_0)$(mm) |
|---|---|---|---|---|
| 1 | 0 | 0 | 1.4142 | 15, 3.6364 |
| 2 | 3 | 0 | 1.3820 | 15, 3.1111 |
| 3 | 6 | 0 | 1.2806 | 15, 1.2525 |
| 4 | 9 | 0 | 1.0909 | 9.6465, 0 |
| 5 | 1.5 | $0.5\sqrt{3}$ | 1.4036 | 15, 3.4747 |
| 6 | 4.5 | $0.5\sqrt{3}$ | 1.3379 | 15, 2.3434 |

| 7  | 7.5 | 0.5√3 | 1.1958 | 14.3939, 0 |
| 8  | 0   | 0.5√3 | 1.4036 | 15, 3.4747 |
| 9  | 3   | 0.5√3 | 1.3711 | 15, 2.9091 |
| 10 | 6   | 0.5√3 | 1.2689 | 15, 1.0505 |
| 11 | 9   | 0.5√3 | 1.0770 | 8.7374, 0  |
| 12 | 1.5 | 1.5√3 | 1.3820 | 15, 3.1111 |
| 13 | 4.5 | 1.5√3 | 1.3153 | 15, 1.8990 |
| 14 | 7.5 | 1.5√3 | 1.1705 | 13.5859, 0 |
| 15 | 0   | 2√3   | 1.3711 | 15, 2.9091 |
| 16 | 3   | 2√3   | 1.3379 | 15, 2.3434 |
| 17 | 6   | 2√3   | 1.2329 | 15, 0.3232 |
| 18 | 9   | 2√3   | 1.0344 | 5.5051, 0  |
| 19 | 1.5 | 2.5√3 | 1.3379 | 15, 2.3434 |
| 20 | 4.5 | 2.5√3 | 1.2689 | 15, 1.0505 |
| 21 | 7.5 | 2.5√3 | 1.1180 | 11.1616, 0 |
| 22 | 0   | 3√3   | 1.3153 | 15, 1.9394 |
| 23 | 3   | 3√3   | 1.2806 | 15, 1.2525 |
| 24 | 6   | 3√3   | 1.1705 | 13.5859, 0 |
| 25 | 1.5 | 3.5√3 | 1.2689 | 15, 1.0101 |
| 26 | 4.5 | 3.5√3 | 1.1958 | 14.3939, 0 |
| 27 | 7.5 | 3.5√3 | 1.0344 | 5.5051, 0  |
| 28 | 0   | 4√3   | 1.2329 | 15, 0.3232 |
| 29 | 3   | 4√3   | 1.1958 | 14.3939, 0 |
| 30 | 6   | 4√3   | 1.0770 | 8.7374, 0  |
| 31 | 1.5 | 4.5√3 | 1.1705 | 13.5859, 0 |
| 32 | 4.5 | 4.5√3 | 1.0909 | 9.6465, 0  |
| 33 | 0   | 5√3   | 1.1180 | 11.1616, 0 |
| 34 | 3   | 5√3   | 1.0770 | 8.7374, 0  |
| 35 | 1.5 | 5.5√3 | 1.0344 | 5.5051, 0  |

Table 1: Geometric Parameters of the Designed Lens Microstructure